\documentstyle[preprint,aps]{revtex}
\begin{document}
\preprint{OZ-93/22, UM-P-93/89,
 OITS 519}
\draft
\title{
CP Violation in Fermion Pair Decays of
Neutral Boson Particles}
\author{Xiao-Gang He$^1$, J. P. Ma$^2$ and Bruce McKellar$^2$ }
\address{
$^1$Institute of Theoretical Science\\
University of Oregon\\
Eugene, OR97403, USA\\
and\\
$^2$Research Center for High Energy Physics\\
School of Physics\\
University of Melbourne \\
Parkville, Vic. 3052 Australia}
\date{October, 1993}
\maketitle
\begin{abstract}
We study CP violation in fermion pair decays of neutral boson particles
with spin 0 or 1.
We study a new asymmetry to
measure CP violation in $\eta, K_L \rightarrow \mu^+\mu^-$ decays
and discuss the
possibility  of measuring it experimentally. For the spin-1 particles
case, we study CP violation in the decays of $J/\psi$ to $SU(3)$ octet baryon
pairs. We show that these decays can be used to put stringent constraints on
the electric dipole moments of $\Lambda$, $\Sigma$ and $\Xi$.

\end{abstract}
\newpage
\section{Introduction}
CP violation has only been observed in the neutral Kaon system\cite{1}.
In order to
isolate the source (or sources) responsible for CP violation, it is important
to find CP violation in other systems. In this paper we study
CP violation in fermion pair decays of
a neutral boson particle, which is a CP eigenstate and has spin-0(S) or spin-1
(V).

a) $S\rightarrow f\bar f$

The most general decay amplitude for S decays into a pair of spin-1/2
particles $f\bar f$ can be parametrized as
\begin{equation}
   M(S\rightarrow f\bar f) =\bar u_f(p_1, s_1) (a_S+i\gamma_5 b_S) v_{\bar f}
(p_2, s_2)\;,
\end{equation}
where $a_S$ and $b_S$ are in general complex numbers. If both $a_S$ and
$b_S$ are nonzero, CP is violated.
One can define a density matrix $R$
for the process $S \rightarrow f\bar f$, where the $f (\bar f )$ is
polarized and the polarization is described by a unit polarization vector
${\bf s}_{1(2)}$ in the $f(\bar f)$ rest frame. With the amplitude in
eq.(1) the CP violating part of the density matrix
in the rest frame of $S$ is given by
\begin{eqnarray}
   R_{CP}= N_f\{{\rm Im}(a_Sb_S^*) {\bf p}\cdot ({\bf s}_1
-{\bf s}_2) -{\rm Re}(a_Sb_S^*) {\bf p}\cdot ({\bf s}_1\times {\bf s}_2)\}\;,
\end{eqnarray}
where $N_f$ is a normalization constant, and
${\bf p} $ is the three momentum of the fermion $f$.
$R_{CP}$ contains all information about CP violation in the decay.
The expectation values
of CP odd and CPT odd observables are
proportional to $Im(a_Sb_S^*)$. This CP violating parameter can be measured
by the asymmetery
\begin{equation}
A(S) = {N_+ -N_-\over N_+ + N_-}\;,
\end{equation}
where $N_{+(-)}$ indicates the decay events
with ${\bf  s}_1\cdot {\bf  p} >0 (<) 0$.
In terms of the parameters in the decay amplitude,
\begin{eqnarray}
A(S) &=& \beta_f {Im(a_S b_S^*)\over \beta_f^2 |a_S|^2 + |b_S|^2}\nonumber\\
&=& {\beta_f^2 M_S Im(a_Sb_S^*)\over 8\pi \Gamma_f}\;,
\end{eqnarray}
where $\Gamma_f$ is the decay width for $S \rightarrow f\bar f$, $\beta_f
=\sqrt{1-4m_f^2/M_S}$, and $m_f$ and $M_S$ are the masses of the fermion
$f(\bar f)$ and the scalar S, respectively. The asymmetry $A$
has been studied extensively for $\eta, K_L \rightarrow \mu^+\mu^-$ decays
\cite{2,3,4,5}.  Here we will
instead study a CP odd and CPT
even observable which can provide additional information about CP violation.
It is related to ${\rm Re}(a_Sb_S^*)$. We construct the
following asymmetry B to probe this CP violating parameter,
\begin{equation}
B(S) = {N^+ - N^-\over N^++N^-}\;,
\end{equation}
where $N^+$ and $N^-$ indicates the decay events with
$({\bf s}_1\times {\bf s}_2)\cdot {\bf  p}>(<)0$.
In terms of the parameters in the decay amplitude, we
have
\begin{eqnarray}
B(S) &=& - {\pi\over 4}\beta_f {Re(a_S b^*_S)\over \beta^2_f|a_S|^2+|b_S|^2}
\nonumber\\
&=&-{\beta_f^2M_SRe(a_Sb^*_S)\over 32\Gamma_f}\;.
\end{eqnarray}

To experimentally measure $A$ and $B$, one must know the polarizations of the
fermions in the final state. The polarizations can be analysed by using certain
decay channels of $f$ and $\bar f$.
Assuming that the polariztions of $f$ and $\bar f$ are
analysed by the decays $f \rightarrow
f'(p_{f'}) + X$ and $\bar f \rightarrow \bar f'(p_{\bar f'}) + \bar X$,
with  density matrices given by

\begin{eqnarray}
\rho_f &=& 1 + \alpha_f {\bf  s}_1\cdot {\bf \hat p}_{f'}\;,\nonumber\\
\rho_{\bar f} &=& 1-\alpha_{\bar f} {\bf  s}_2\cdot {\bf \hat p}
_{\bar f'}\;,
\end{eqnarray}
where $\alpha_{f(\bar f)}$ are constants, and the hat  on the momentum
indicates the unit vector in the direction of the momentum.
Using this information, we define a more convenient asymmetry,
\begin{eqnarray}
\tilde A(S) &=&  {\tilde N_+ - \tilde N_-\over \tilde N_+ + \tilde N_-}
= \alpha_f A(S)\nonumber\\
\tilde B(S) &=& {\tilde N^+ - \tilde N^-\over \tilde N^+ + \tilde N^-}
=-\alpha_f\alpha_{\bar f}B(S)\;,
\end{eqnarray}
where $\tilde N_{+(-)}$
 and $\tilde N^{+(-)}$ indicate events with
${\bf \hat p}_{f'}\cdot {\bf \hat p} >(<)0$ and
$({\bf \hat p}_{f'}\times {\bf \hat p}_{\bar f'})
\cdot {\bf \hat p}_f>(<)0$, respectively.

b) $V\rightarrow f\bar f$

The most general decay amplitude for this decay can be parametrized as
\begin{eqnarray}
M(V\rightarrow f\bar f)
&=& \varepsilon^{\mu}\bar u_f(p_1)[\gamma_{\mu}(a + b\gamma_5) +
(p_{1\mu} -p_{2\mu})(c + id\gamma_5)]v_{\bar f} (p_2)\; ,
\end{eqnarray}
where $\varepsilon^\mu$ is the polarization of  V and in its
rest frame $\varepsilon_\mu = (0, \vec {\bf \varepsilon})$. If CP is conserved,
$d = 0$. The constants
$a,\; b,\;c$ and $d$ are in general complex numbers.

The density matrix for this decay
in the rest frame of V, up to a normalization constant, is given by
\begin{eqnarray}
R_{ij} &=&
[\bar u_{\Lambda}(p_1,{\bf s}_1)[\gamma_{i}(a + b\gamma_5) +
(p_{1i} -p_{2i})(c + id\gamma_5)]v_{\bar {\Lambda}} (p_2,{\bf s}_2)\nonumber\\
&\times& \bar v_{\bar \Lambda}(p_2,{\bf s}_2)[\gamma_{j}
(a^* + b^*\gamma_5) +
(p_{1j} -p_{2j})(c^* + id^*\gamma_5)]u_{\Lambda} (p_1,{\bf s}_1)]\; ,
\end{eqnarray}
where $i$ and $j$ label three-vector components.

The CP violating part of this density matrix is
given by
\begin{eqnarray}
R_{ij}&= &r_{ij} + r_{ji}^*\;,\nonumber\\
r_{ij}&= &i2ad^* p_j\{{M_V^2\over 2} ( {\bf s}_1- {\bf s}_2)_i
- {2M\over {M_V+2m_f}} ({\bf s}_1-{\bf s}_2)\cdot{\bf p}  p_i\nonumber\\
&+&imM_V({\bf s}_1\times {\bf s}_2)_i + i{2M_V\over {M_V+2m_f}}
({\bf s}_1\cdot{\bf p} ({\bf p}\times {\bf s}_2)_i - {\bf s}_2\cdot{\bf p}
({\bf p}\times {\bf s}_1)_i)\}\nonumber\\
&+&2ibd^*M_V p_j\{ s_{2i}{\bf s}_1\cdot{\bf p} -  s_{1i}
{\bf s}_2\cdot{\bf p}
+i ({\bf p}\times ({\bf s}_1 -{\bf s}_2))_j\}\\
&+&4icd^*M_V p_i p_j \{-({\bf s}_1 - {\bf s}_2)\cdot {\bf p}
+i({\bf s}_1\times {\bf s}_2)\cdot{\bf p}\}\nonumber\;,
\end{eqnarray}
where $M_V$ the mass of $V$. In general, $V$ is produced with polarization,
and the polarization depends on how V is produced and is different in
different productions. However, we can construct
a similar asymmetry as for the spin-0 decay case,
which is independent of the polarization due to rotation invariance, to probe
the CP violating parameters\cite{6}, $Re(da^*)$, $Re(dc^*)$,
\begin{eqnarray}
 B(V) &=& {N^+-N^-\over N^+ + N^-}\nonumber\\
&=&-{\beta^2M_V\over 96\Gamma_f} (2m_fRe(da^*)+(M^2_V-4m_f^2)
Re(dc^*))\;,\nonumber\\
\tilde B(V) &=& {\tilde N^+-\tilde N^-\over \tilde N^+ + \tilde N^-}\nonumber\\
&=&-\alpha_f\alpha_{\bar f}B(V)\;.
\end{eqnarray}
Here $\beta = \sqrt{1-4m_f^2/M_V^2}$.

\section{CP violation in $s\rightarrow f\bar f$}
In this section we study the asymmetry $B$ for
$\eta, K_L \rightarrow \mu^+\mu^-$ decays.
CP violating tests in these systems have been studied before\cite{2,3}.
All of them
concentrated on the asymmetry $A$. Here we show that the asymmetry $B$ is also
a
good quantity to study CP violation. It reveals information not contained in
$A$.

\noindent
$\eta \rightarrow \mu^+\mu^-$.

Because $\eta$ is a pseudo-scalar, if CP is conserved, $a_\eta = 0$. The CP
violating contributions are expected to be small. We can use the decay width
to determine $b_\eta$. $Imb_\eta$ is
determined from $\eta \rightarrow \gamma\gamma \rightarrow \mu^+\mu^-$ with the
two intermediate photons on-shell. Using experimental data for $\eta
\rightarrow
\gamma\gamma$\cite{7}, one obtains
\begin{eqnarray}
|\mbox{Im}b_\eta| &=& {\alpha_{em}\over 4\beta}{m_\mu\over m_\eta}
\mbox{ln} {1+\beta\over 1-\beta}
[64\pi\Gamma(\eta\rightarrow 2\gamma)/m_\eta]^{1/2}\nonumber\\
&=& 1.59\times 10^{-5}\;.
\end{eqnarray}
This amplitude is close to the experimental amplitude determined from
$Br(\eta \rightarrow \mu^+ \mu^-) = (5\pm 1)\times 10^{-6}$\cite{8}.
The real part of
the amplitude is $|\mbox{Re}b|
\approx 0.7\times 10^{-5}$. Using these numbers we find
\begin{eqnarray}
|B(\eta)| = 2\times 10^4 |\mbox{Re}a_\eta + 2.3 \mbox{Im}a_\eta|\;.
\end{eqnarray}
Here we have assumed that $\mbox{Re}b_\eta$ and $\mbox{Im}b_\eta$
have the same sign.
The asymmetry $A$ is $5.8\times 10^4(\mbox{Re}a_\eta - 0.44\mbox{Im}a_\eta)$.

The parameter $a_\eta$
 is model dependent. In many model $a_\eta$ is very small\cite{2,3}. In
lepton-quark models, the constraint on $a_\eta$
is from the neutron electric dipole
moment. If one assumes no cancellations among different contributions to
the neutron electric dipole moment, $a_\eta$ is constrained to be less than
$2\times
10^{-9}$. However, if one allows cancellations between different
contributions, it is possible
to have relatively
large $a_\eta$\cite{3}. The asymmetry $B$ can reach $10^{-3}$ or even
larger. The polarization of
the muons from $\eta\rightarrow \mu^+\mu^-$ can be analysed by $\mu
\rightarrow e \bar \nu_e \nu_\mu$. In
this case $\alpha_e = 1/3$. The $\eta$ factory at SACLAY can reach a
sensitivity
for $A$ and $B$ at the level a few \% in the near future\cite{9}.
It may provide
interesting information about CP violation.

\noindent
$K_L\rightarrow \mu^+\mu^-$.

Using data from $K_L \rightarrow 2\gamma$ and
$K_L \rightarrow \mu^+\mu^-$, we obtain, $|\mbox{Im}b_K|
= 2\times 10^{-12}$, and
$|\mbox{Re}b_K| = (0.14 \pm 0.16)\times 10^{-12}$.
The contributions to the asymmetries
$A$ and $B$ from direct CP violation are given
by
\begin{eqnarray}
|A(K_L)| = 3.6\times |\mbox{Re}a_K - 0.07 \mbox{Im}a_K|\;,\nonumber\\
|B(K_L)| =2 \times 10^{11} |0.1\mbox{Re}a_K + 1.4 \mbox{Im}a_K|\;.
\end{eqnarray}
We have used the central values for $Reb_K$ and $Imb_K$.  In the above analysis
we have assumed that $K_L$ is a
pure CP egeinstate. This is, however, not the whole story.
$K_L$ is not a pure CP
eigenstate,
\begin{equation}
K_L = {1\over \sqrt{1 + |\epsilon|^2}}(|K_2> + \epsilon |K_1>)\;,
\end{equation}
where $CP|K_2> = -|K_2>$, $CP|K_1> = |K_1>$
and the mixing parameter $\epsilon$ is
measured to be $2.27\times 10^{-3}e^{i\pi/4}$. The asymmetries  $A$ and
$B$ are related to $\mbox{Im}(b_2(a_2+\epsilon b_1))$ and
$\mbox{Re}(b_2(a_2 + \epsilon b_1))$, respectively. Here $a_i$ and $b_i$
are the amplitudes for $K_i \rightarrow
\mu^+\mu^-$. The parameter $b_1$ is not zero. Using the values for the real and
 the imaginary parts of $b_1$ determined in Ref.\cite{10},  and set $a_2=0$, we
obtain
\begin{eqnarray}
B(K_L)|_{a_2=0} \approx 0.3\times 10^{-3}\;.
\end{eqnarray}
The CP odd and CPT odd asymmetry $A(K_L)|_{a_2} \approx 10^{-3}$.
If experiments measure larger value for $A$ and $B$,
there must be new physics due to large $a_2$. In many models the paramter $a_2$
is predicted to be very small. However there are models which can produce
large $a_2$\cite{4}.
$B$ can be as large as $10^{-2}$.
A Kaon factory may be able to see CP violation.

In both $\eta\rightarrow \mu^+\mu^-$ and $K_L \rightarrow \mu^+\mu^-$
decays, the experimental sensitivities for the asymmetry $A$ is slightly better
than for $B$. If $Rea$ is larger than $Ima$, the asymmetry $A$ is a better
quantity to measure. However, if it turns out that $\mbox{Im}a > \mbox{Re}a$,
the asymmetry $B$ is the better one.

The decay $B_d (B_s) \rightarrow \mu^+ \mu^-$
can be used to study CP violation also.
However in the standard model, the branching ratios for these decays are very
small\cite{11}.
If the standard model prediction is correct, it is very difficult to test
CP violation  using these decay modes.
One, of cause, should keep in mind
that should this decay be discovered with a branching ratio larger than
the standard model predictions,
there must be new physics and large CP violation may be observed.
The same comments apply to $D^0\rightarrow \mu^+\mu^-$. The same
analysis can also be carried out for Higgs particle decays\cite{12}.

\section{$V\rightarrow f \bar f$}

The decays $V \rightarrow f\bar f$
provide new tests for CP violation. In a previous paper we have
studied a particular case, $J/\psi \rightarrow \Lambda\bar \Lambda$, and shown
that this is a good place to look for CP violation\cite{6}.
In this section we
will carry out a more detailed analysis by including more decay channels.

\noindent
$J/\psi \rightarrow B_8\bar B_8$.

The branching ratio for $J/\psi$ decays into baryon pairs $B_8$ and $\bar B_8$
of the $SU(3)$ octet is typically $10^{-3}$.
With enough $J/\psi$ decay events, we  may obtain useful information
about CP violation.  In Eq.(9), the $b$-term is a P violating amplitude and is
expected to be significantly smaller than the P conserving $a$- and $c$-
amplitudes.
We will therefore neglect the contribution to the branching ratio from $b$.
The relative strength of the amplitude $a$ and $c$ can be determined by
studying angular correlations between the polarization of
$J/\psi$ and the direction of $B_8$ momentum. Due to large
experimental uncertainties associated with the constants which determine the
angular distribution, $a$ and $c$ can not be reliably determined separately at
present\cite{7}. In our numerical
estimates we will consider two cases where the decay ampiltudes are
dominated by
1) the $a$-term, and 2) the $c$-term, respectively. Assuming $a$ and $b$
are real and
using the experimental branching ratios compiled by the particle data group, we
obtain numerical values for the asymmetry $B$. The results are given in Table
I.

The CP violating $d$-term can receive contributions
from different sources, the electric dipole moment, the CP violating Z
-$B_8$ coupling,  etc. In the following we estimate the
contribution from the electric dipole moment $d_{B_8}$ of $B_8$.
Here $d_{B_8}$ is defined as
\begin{equation}
L_{dipole} = i{d_{B_8} \over 2} \bar B_8 \sigma_{\mu \nu}\gamma_5
B_8 F^{\mu \nu}\;,
\end{equation}
where $F^{\mu \nu}$ is the field strength of the electromagnetic field.
Exchanging a photon between $B_8$ and a c-quark, we have the CP violating
c-$B_8$ interaction
\begin{equation}
L_{c-\Lambda} = -{2\over 3M^2} e d_{\Lambda}(p_1^\mu - p_2^\mu)\bar c
\gamma_\mu c \bar B_8 i\gamma_5B_8\;.
\end{equation}
{}From this we obtain
\begin{equation}
d =- {2\over 3}  {g_V\over M_{J/\psi}^2}e d_{B_8}\;.
\end{equation}
Here we have used the parametrization,
$<0|\bar c \gamma_\mu c|J/\psi> = \varepsilon_\mu g_V$. The value $|g_V|$ is
determined to be 1.25 GeV$^2$ from $J/\psi \rightarrow \mu^+ \mu^-$. There are
additional contributions to $d$, for example, exchanging a photon between the
final $B_8$ and $\bar B_8$. We have checked several contributions of this type
and find them to be small if electric dipole moment is the only source of CP
violation. Using
the above information, we can express the asymmetry $B$ in
terms of the electric
dipole moment of the baryons. The asymmetry $B$ can be used to put constraints
on the electric dipole moment. In Table II we give the asymmetry $B$ in terms
of the electric dipole moment of $B_8$. Note that because photons are
off-shell, $d_{B_8}$ is measured at $q^2 = M_{J/\psi}^2$ from the measurement
of $B$. If we assume that the extrapolation follow the same $q^2$ dependence as
the magnetic dipole moment of $B_8$, $d_{B_8}(q^2=M_{J/\psi}^2)$ is
smaller than $d_{B_8}(q^2=0)$. However, the $q^2$ dependence of the electric
dipole moment may be completely different from the magnetic dipole moment. It
is possible that $d_{B_8}$ does not change very much from $q^2=0$ to
$q^2=M_{J/\psi}^2$.

The polarizations  of $B_8$ and $\bar B_8$ in $J/\psi \rightarrow
B_8\bar B_8$, can be analysed by certain decay channels of $B_8$ and
$\bar B_8$. There are many decay channels available to carry out such
analysis\cite{7}.
The neutron polarization can be analysed by $n \rightarrow p e\bar \nu_e$.
The
proton polarization may be analysed by rescattering. It
may be difficulty to carry out such analysis. The polarization of $\Lambda$ can
be analysed by, for example, $\Lambda \rightarrow p \pi^-$. This decay mode has
a large branching ratio (64\%) and a large
$\alpha_\Lambda$ (0.642).
The polarization of $\Sigma$ can also be analysed. For
$\Sigma^-$, one can use $\Sigma^- \rightarrow n\pi^-$. This decay mode has
large
branching ratio (99.85\%) with $\alpha_{\Sigma^-} = -0.068$. The
polarization of $\Sigma^0$ can be analysed by $\Sigma^0 \rightarrow \Lambda
\gamma$.
This is the dominant decay channel for $\Sigma^0$ (100\%).
The polarization of $\Sigma^+$ can be
analysed by $\Sigma^+ \rightarrow p\pi^0$. The branching ratio is 51.6\% and
has
a large value for $\alpha_{\Sigma^+}$ (-0.98). The polarization of $\Xi^0$
can be analysed by $\Xi^0 \rightarrow \Lambda \pi^0$. This is the main decay
channel (100\%) and the parameter $\alpha_{\Xi^0} = -0.411$. The polarization
of $\Xi^-$ can be analysed by $\Xi^-\rightarrow \Lambda\pi^-$, again this
decay mode is the dominant one (100\%) and has a large vaule for $\alpha_{
\Xi^-}$ (-0.456).

The asymmetry $B$ may not be useful in providing upper bounds for the
electric dipole moment for neutron and proton because their electric dipole
moments have been constrained to be very small, $d_n < 1.2\times 10^{-26} ecm$
\cite{13}
and $d_p < 10^{-22}ecm$\cite{14}.
However useful information about the electric dipole moments
for $\Lambda$, $\Sigma$ and $\Xi$ can be extracted.
The experimental upper bound on $d_{\Lambda}$ is $1.5 \times 10^{-16}
\rm{ecm}$\cite{7}. There are constraints on the strange quark electric
dipole moment and colour dipole moment from the neutron electric dipole
moment $d_n$ , which follow if one assumes that the contributions to
$d_n$ do not cancel against each other\cite{15}. It is
possible that cancellations do occur for $d_n$ but not $d_\Lambda$ and the
constraints from $d_n$ do not necessarily lead to strong constraints on
$d_\Lambda$. Alternative experimental approaches to $d_\Lambda$, such as that
presented here, should therefore be pursued.
If $d_\Lambda$ indeed has a value  close to its experimental upper bound,
the asymmetry $B$ can be as large as $O(10^{-2})$. Using $\Lambda
\rightarrow p\pi^-$ to analyse the polarization, we can obtain $\tilde B$
as large as $10^{-2}$.  With $10^{7}$
$J/\psi$, it is already possible to obtain some interesting results. This
experiment can be performed with the Beijing $e^+$ $e^-$ machine.
If $10^{9}\; J/\psi$ can be produced, one can improve the  upper bound on
$d_\Lambda$ by an order of magnitude.
This can be achieved in future $J/\psi$ factories.
There is not much information about the electric dipole moment of $\Sigma$
and $\Xi$. The observable $B$ can thus
be used to put upper bound on the electric
dipole moments of $\Sigma$ and $\Xi$. With $10^9$ $J/\psi$ decays, the
sensitivity for the elelctric dipole moment is typically $10^{-17}ecm$.

Our analysis can also be used for $J/\psi \rightarrow
l^+l^-$.
Assuming that the d-term is mainly due to
the electric dipole moment $d_l$ of the lepton, we have
\begin{eqnarray}
B &=& {d_l \over e}{\pi \over 4} m_l{\sqrt{1 - 4m_l^2/M^2}\over {1+2m_l/M}}\;,
\end{eqnarray}
where $m_l$ is the lepton mass.  For $J/\psi
\rightarrow \mu^+ \mu^-$, we have, $B = 4\times 10^{-7}(d_\mu/
{10^{-19}\;\rm{ecm}})$ which may be too small to be measured experimentally.

\noindent
$\Upsilon \rightarrow f\bar f$.

In principle the asymmetry $B$ can be used to probe CP violation in
$\Upsilon \rightarrow B_8\bar B_8$.
For these decays the branching ratios are not measured yet. They are
smaller than the branching ratios for $J/\psi \rightarrow B_8\bar B_8$,
e.g. $Br(\Upsilon \rightarrow p\bar p) <9\times 10^{-4}$.
In order to reach the same sensitivity for CP violating parameters as
for $J/\psi \rightarrow B_8\bar B_8$, more
$\Upsilon$ events are needed. It may not be practical to study CP violation
using these decay modes.

 It may be possible to observe CP violation in $\Upsilon \rightarrow l\bar l$.
One particular interesting decay mode is $\Upsilon \rightarrow \tau^+\tau^-$.
The tauon polarization can be analysed by the decays $\tau \rightarrow \pi
\nu\;, 2\pi\nu\;, 3\pi\nu\;, e\nu\bar\nu$ and $\mu \nu\bar\nu$.
 It has been shown that these decay channels provide reasonable sensitivity
for tauon polarization analysis\cite{16}.
Assuming the electric dipole moment of the tauon is the source for CP violation
in this decay, $B =
7\times 10^{-3} d_\tau / (10^{-16}\; \rm{ecm})$.
The experimetal upper bound on $d_\tau$ is $1.6\times 10^{-16}\; \rm{ecm}$,
 so the asymmetry $B$ can be as large as $10^{-2}$.
Values of $d_\tau$ as large as $10^{-16} \rm{ecm}$ can be obtained in model
calculations. The leptoquark model is one of them. In this model
there is
a scalar which can couple to leptons and quarks. The couplings of the
leptoquark scalar
to the third generation are weakly constrained\cite{17}.
It is possible to generate
a large $d_{\tau}$ by exchanging a leptoquark at the one loop level.

Similar  experiments can be carried out for other systems,
for example, $\rho\;, \phi \rightarrow \mu^+\mu^-$ and
$Z\rightarrow l\bar l$\cite{18}..
In particular the $\phi$ factory may provide useful information about CP
violation in $\phi \rightarrow \mu^+\mu^-$.

\section{conclusion}
We studied CP violation in fermion pair decays of spin-0 and spin-1 particles
using a CP odd and CPT even observable. The asymmetry $B$ studied in this
paper provides another test  for CP violation in
$\eta, K_L \rightarrow \mu^+\mu^-$ decays. This asymmetry can reveal new
information which is not contained in the asymmetry $A$ studied previously in
the liturature.
For spin-1 particle
case, we studied CP violation in
the decays of $J/\psi$ to the $SU(3)$ octet baryon
pairs. We showed that these decays can be used to put
stringent constraints on
the electric dipole moments of $\Lambda$, $\Sigma$ and $\Xi$. Using the
$J/\psi$
events accumulated at the Beijing $e^+e^-$ collider, one may already obtain
interesting information about CP violation. We encourage our experimental
collegaues to carry out such analysis.

\acknowledgments
XGH would like to thank Pakvasa for his hospitality at the University of Hawaii
where part of this work was carried out. He also would like to thank Deshpande,
Pakvasa and Tata for useful discussions. This work is supported in part by the
Australian Research Council and by the
U.S. Department of Energy under contract DE FG06-85ER40224.

\begin{table}
\caption{The asymmetry B.}
\begin{tabular}{|c|c|c|}
Decay Mode & $a$-term Dominates (in unit d($GeV$)) & $b$-term Dominates (in
unit d($GeV$)\\ \hline
$n\bar n$& $3.5\times 10^{2}$&$9.0\times 10^2$\\
$p\bar p$& $3.2\times 10^2$&$8.2\times 10^2$\\
$\Lambda \bar \Lambda$& $3.8\times 10^2$&$8.4\times 10^2$\\
$\Sigma\bar\Sigma$ &$3.7\times 10^2$ &$7.6\times 10^2$\\
$\Xi\bar\Xi$&$3.1\times 10^2$&$6.1\times 10^2$ \\
\end{tabular}
\label{table1}
\end{table}

\begin{table}
\caption{The asymmetry $B$ in terms of the electric dipole moment of $B_8$.}
\begin{tabular}{|c|c|c|}
Decay Mode& a-term Dominates (in unit $10^{14}/ecm)$&
b-term Dominates (in unit $10^{14}/ecm)$\\ \hline
$n\bar n$&$1.38d_n$&$3.5d_n$\\
$p\bar p$&$1.25d_p$&$3.2d_p$\\
$\Lambda\bar \Lambda$&$1.48d_\Lambda$&$3.3d_\Lambda$\\
$\Sigma\bar\Sigma$&$1.46d_\Sigma$&$2.9d_\Sigma$\\
$\Xi\bar\Xi$&$1.22d_\Xi$&$2.4d_\Xi$\\
\end{tabular}
\label{table2}
\end{table}
\end{document}